\documentclass[twoside,fleqn]{article}
\usepackage{espcrc2}

\newcommand{\AmS}{{\protect\the\textfont2
  A\kern-.1667em\lower.5ex\hbox{M}\kern-.125emS}}

\title{'t Hooft Tensor for generic Gauge Groups}

\author{A. Di Giacomo\thanks{Presented the talk}\address{Dip. di Fisica Universita' di Pisa and INFN-Sezione di Pisa\\
        Largo B. Pontecorvo 3, 56127 Pisa, Italy}
        L. Lepori\address {International School for Advanced Studies (SISSA) and INFN-Sezione di Trieste\\
        Via Beirut 2-4, 34014 Trieste, Italy}
        F.Pucci\address{Dip. Fisica Universita' di Firenze and INFN-Sezione di Firenze\\
        Via G. Sansone 1, 50019 Sesto Fiorentino, Italy} }

\begin{document}
\begin{abstract}
We study monopoles in gauge theories with generic gauge group.
Magnetic  charges are in one-to-one correspondence with the second
homotopy classes at spatial infinity (${\Pi}_2$), which are
therefore identified by the 't Hooft tensor. We determine the 't
Hooft tensor in the general case. These issues are relevant to the
understanding of Color Confinement.
 \end{abstract}

\maketitle

\section{INTRODUCTION}

No quark has ever been observed in Nature.

The ratio of the abundance of quarks in ordinary matter $n_q$ to
the abundance of protons $n_p$ has an experimental upper limit
${{n_q} \over{ n_p}} \le 10^{-27}$, to be compared to the
expectation in the Standard Cosmological Model \cite{okun} ${{n_q}
\over {n_p} }\approx 10^{-12}$. The inhibition factor which
quantifies confinement is $ \approx 10^{-15} $.\\
A similar factor
limits the production of quarks in high energy reactions. The
cross section for inclusive production of quarks plus antiquarks
has an experimental upper limit $\sigma_q \equiv \sigma (p+p
\rightarrow q(\bar q)+ X) \le 10^{-40} cm^2$, to be compared to
the perturbative estimate $\sigma_q \approx {\sigma_{TOT} }
\approx {10^{-25} cm^2}$  Again an inhibition factor of ${\approx
10^{-15} }$.\\
 The natural explanation is that  $n_q =0 $
and $\sigma_q = 0$  protected by some symmetry: this is similar
to what happens in ordinary superconductivity, where the
resistivity is constrained by experiment to be a very small
fraction of that of the normal metal. The resistivity of the
superconductor is strictly zero, due to the Higgs breaking of
electric charge $U(1)$ symmetry which is restored in the normal
phase.\\
 If this is true the deconfining transition is a change of
symmetry, i.e. an order-disorder transition, and can not be a
crossover. In particular an order parameter must exist which
unambiguously defines confinement versus deconfinement.\\
Two main questions follow immediately:

 a) What is the symmetry related to confinement?  Color is an exact symmetry, both in confined and in deconfined phase, and therefore can not be our symmetry. We need an extra  symmetry besides color.

 b) There is no direct experimental observation yet of the deconfining transition, but it has been observed in lattice simulations. Are observations on the  lattice compatible with an order disorder transition?

 The question  b) is still open \cite{owen}\cite{DDP}\cite{CDDP} and will not be discussed here.
 We shall instead address here the question  a).

 In the absence of quarks the action is blind to the center $Z_3$ of the gauge group. The theory  can be however
formulated in terms of  $3 \times 3 $ matrices of the fundamental
representation of the gauge group, to allow the introduction of
static external quarks. $Z_3$ is then an extra symmetry, and the
Polyakov line  an order parameter,  detecting confinement and
deconfinement.  In presence of dynamical quarks,
 as in Nature, $Z_3$ is explicitly broken and therefore can not be the symmetry  we look for.
An extra symmetry can only be provided by dual degrees of freedom,
i.e. by infrared modes with non trivial spatial homotopy, or by
global properties of the field configurations, which exist besides
local gauge symmetry.

The natural excitations are vortices in (2+1)-dimensional QCD,
where the surface at spatial infinity is a circle and the homotopy
$\Pi_1$, they are monopoles in the physically realistic
case of QCD in $(3+1)$ dimensions, where the surface at spatial
infinity is that of a sphere and the  homotopy  $\Pi_2$.

\section{MONOPOLES}
The prototype monopole configuration in gauge theories is that of
ref\cite{'tH}\cite{Pol}. It is a soliton solution in the Higgs
broken phase of an SO(3) gauge theory coupled to a Higgs field in
the adjoint representation.
\begin{equation}
L= -{1\over 4} {\vec G_{\mu \nu}} {\vec G_{\mu \nu}} + (D_{\mu} \vec\Phi)^2 - V(\Phi^2)
\end{equation}
In the "hedgehog" gauge the soliton solution is
\begin{equation}
\phi^a = f(r) {r^a \over r}\\
A^a_i = g(r) \epsilon _{aij}{r^i\over r^2}
\end{equation}
$f(r)$ and $g(r)$ are $\approx 1$ outside a radius which is
determined by the Higgs $vev $  $ \langle\Phi\rangle$. $ \phi^a $
has been normalized to be $1$ at  $x\to \infty$. The solution is a
non trivial mapping of the 2-dim sphere at infinity $S_2$ onto
$SO(3)/U(1)$. Going to the unitary gauge where $ \vec \phi  = ( 0
, 0 ,  1)$, a line-like singularity appears starting from $\vec x
= 0$. At large distances the abelian field strength of the residual $U(1)$ gauge symmetry in the unitary gauge
  \begin{equation}
  F^3_{\mu\nu} = \partial _{\mu} A^3_{\nu} - \partial _{\nu} A^3_{\mu}
  \end{equation}
   is  for the soliton solutions
\begin{displaymath}
 E_i\equiv   F^3_{0i}  =0\end{displaymath}\begin{equation} \vec H = {1\over g}{{\vec r} \over {4\pi r^3} } + \texttt{Dirac - string}\end{equation}
with $H_i = {1\over 2} \epsilon _{ijk} F^3_{jk}$.
 If the string is invisible, as happens in a compact formulation like lattice, there is a violation of Bianchi identities
 \begin{equation}
 \vec \nabla \vec H = {1\over g} \delta^3(\vec x)
 \end{equation}
 More formally, one can define a gauge invariant tensor\cite{'tH}, the 't Hooft tensor  $F_{\mu\nu}$, which is equal
  to the abelian field strength $F^3_{\mu\nu} $  of eq(3) in the unitary gauge.
  \begin{equation}
  F_{\mu\nu} \equiv  \vec \Phi \vec G_{\mu\nu} - {1\over g} \vec \Phi (\vec {D_{\mu}\Phi} \wedge \vec {D_{\nu}\Phi})
  \end{equation}
  One can show that
  \begin{equation}
  F_{\mu\nu}= \partial_{\mu} (\vec \phi \vec A_{\nu}) - \partial_{\nu} (\vec \phi \vec A_{\mu}) -{1\over g} \vec \phi (\partial_{\mu}\vec \phi \wedge \partial_{\nu}\vec \phi)
  \end{equation}
  which is exactly $F^3_{\mu\nu}$ of eq(3) in the unitary gauge where $\partial _{\mu} \vec \phi =0$.

If we denote by $\tilde F_{\mu\nu} = {1\over 2} \epsilon
_{\mu\nu\rho\sigma} F_{\rho\sigma}$ the dual of $F_{\mu\nu}$, we
can define a magnetic current $j_{\nu}$ as
\begin{equation}
j_{\nu} \equiv  \partial_{\mu} \tilde F_{\mu\nu}
\end{equation}
A non zero $j_{\nu} $ indicates violation of Bianchi identities,
i.e. presence of magnetic charges. Whatever the lagrangian is, the
antisymmetry of  $F_{\mu\nu}$ implies
   \begin{equation}
   \partial_{\nu} j_{\nu} = 0
   \end{equation}
This defines the dual symmetry, which is nothing but conservation
of magnetic charge. Formally  the values of the charge are the
elements of the homotopy group $\Pi_2( SO(3)/U(1))$ which is
easily computed to be $Z/ Z_2$. Only even valued magnetic charges
are allowed.

   A real Higgs breaking is only needed if one wants monopoles as solitons. The field configurations can in any case be classified by their homotopy and the dual current can be defined anyhow.
    Also the Higgs field is not needed: any operator $\Phi$ in the adjoint representation can be used
    to define $F_{\mu\nu}$ and the dual symmetry. Monopole singularities will be located at the zeroes
    of $\Phi$  which are a non dense set \cite{DD} and the particular choice of $\Phi$ becomes irrelevant
    if one defines the field theory in space-time with a discrete set of singularities excluded \cite{Gukov}.
     In particular any operator $\mu$ which creates a monopole\cite{d}\cite{dd}\cite{ddd} adds one extra singularity to field configurations, i.e. a monopole, for whatever choice of $\Phi$. Its $vev$  $\langle \mu \rangle$
     detects dual superconductivity of the vacuum and color confinement, since
     $\langle \mu \rangle = 0 $ when the vacuum state has definite magnetic charge, i.e. when
     the $U(1)$ symmetry is realized $a$ la Wigner (deconfined phase). If, instead,  $\langle \mu \rangle \neq  0 $
     the vacuum is a Bogoliubov superposition  of states with different magnetic
      charge, i.e. a dual superconductor. Since $\mu$ creates a monopole for whatever choice of the
       operator $\Phi$ the statement is independent on it.
\section{ EXTENSION TO GENERIC GAUGE GROUPS \cite{dlp}}
The gauge group of strong interactions in Nature is color $SU(3)$.
However, to get insight into the mechanism of confinement,
theories with different gauge groups can be studied. For example, to clarify the role played by center vortices a study of confinement for gauge groups which have trivial center, and hence no vortices can be useful \cite{h}: this is the motivation for what follows. 
To identify field configurations with
non trivial $\Pi_2$, one needs an $SU(2)$ subgroup of the gauge
group  $G$ to be broken to $U(1)$, e.g. to its third generator.
The $SU(2)$ subgroups of $G$ are easily visualized by looking at
the Lie algebra $g$. $g$ is spanned by the Cartan commuting
operators $H_i$ $(i= 1, ..,r)$, with $r$ the rank of the group,
and by  $E_{\pm \vec \alpha}$, with $\vec \alpha$ the roots of $g$. In
the standard notation
\begin{eqnarray*}
&& [H_i , H_j] =  0\\
&& [H_i, E_{\pm \vec \alpha} ] = \pm\, \vec \alpha\, E_{\pm \vec \alpha} \\
&& [E_{\vec \alpha} ,  E_{\vec \beta}]  =  N_{\vec \alpha \vec \beta} E_{\vec \alpha + \vec \beta}\\
&& [E_{\vec \alpha} ,  E_{-\vec \alpha}]  =  (\vec \alpha\, \vec H)
\end{eqnarray*}
To each root ${\vec \alpha}$ an $SU(2)$ subgroup is associated as can be seen operating a trivial renormalization of the generators:
\begin{eqnarray}
T^{\alpha}_{\pm} &\equiv&  \sqrt{2\over(\vec \alpha \vec \alpha)} E_\pm \vec \alpha\\
T^{\alpha}_3  &\equiv&  {(\vec \alpha \vec H) \over (\vec \alpha \vec \alpha)}
\end{eqnarray}
obeying to the commutation rules:
\begin{eqnarray}
&& [ T^{\alpha}_3 , T^{\alpha}_{\pm}] = \pm T^{\alpha}_{\pm}\\
&& [T^{\alpha}_{+} , T^{\alpha}_{-} ] = 2  T^{\alpha}_3
\end{eqnarray}
A root $\vec \alpha$ is called positive if its first non zero
component is positive: either $\vec \alpha$ or $-\vec \alpha$ is a
positive root. A positive root is called simple if it cannot be
written as a sum of two positive roots.  Any positive root can be
made simple by a suitable (Weyl) transformation of the group.
 Without loss of generality, we can then only consider the monopoles associated to simple roots.
 Simple roots are represented by little circles in the Dynkin diagram of the algebra.\\
 Let $\phi ^a$ be the fundamental weight corresponding to the simple root $\vec a$. Since $\phi ^a$
 commutes with the generators corresponding to all the other simple roots different from $\vec a$,
 the invariance group of $\phi^a$ is a group having as Dynkin diagram the diagram obtained by
 erasing the little circle corresponding to the root $\vec a$  times the $U(1)$ generated by $\phi ^a$ itself \cite{dlp}.

 For gauge group $SU(N)$ there are $(N-1)$ simple roots and the Dynkin diagram has the form
 \begin{center}$\bigcirc$---$\bigcirc$---$\bigcirc$---\, $\ldots$\, $\bigcirc$---$\bigcirc$ \end{center}
 with $(N-1)$ simple roots of equal length $\vec a^i$,  $i= 1, \dots ,N-1$. By erasing the
 i-th simple root one obtains the Lie algebra of the invariance group, which is that
  of $SU(i) \otimes SU(N-i) \otimes U(1)$.  In considering global features the knowledge of Lie algebra is not
 sufficient to identify the real invariance group. For example, in the case of $SU(N)$ the
elements of the center of $SU(i)$ and $SU(N-i)$ can also be
elements of the $U(1)$ group generated by  $\phi ^a$  so that,
when embedding in $SU(N)$, there is a non trivial kernel, which is
composed by the set of elements of  $SU(i) \otimes SU(N-i) \otimes
U(1)$ which are mapped to the identity of $SU(N)$.  By carefully
taking this into account one realizes that
 topologies are in one-to-one correspondence with magnetic charges. This is not only true for $SU(N)$
 but also for any compact simple group, including the exceptional groups \cite{dlp}.

For each simple root $\vec a$ a 't Hooft tensor $F^a_{\mu \nu}$
can be defined as the abelian field strength in the unitary gauge
and with it also a conserved current
 \begin{equation}
 j^a_{\nu}  =  \partial_{\mu} {\widetilde{F}^a}_{\mu \nu}
 \end{equation}
 \begin{equation}
 \partial_{\nu} j^a_{\nu} =0
 \end{equation}
 The corresponding conserved charge is the magnetic charge $Q^a$.
In the deconfined phase the operator $Q^a$ is well defined and
magnetic charge is superselected. In the confined phase, instead,
the corresponding gauge symmetry is broken a la Higgs, and
the vacuum is a superposition of states with different magnetic
charge.
   A set of operators  $\mu^a$ can be defined, which carry non zero $Q^a$ magnetic charge, and
   in terms of them $r$ order parameters $\langle \mu^a \rangle$ for detecting confinement.
    In the deconfined phase magnetic charges are super-selected and $\langle \mu^a \rangle=0$.
    If $\langle \mu^a \rangle \neq 0$ the symmetry is Higgs broken, the vacuum is a dual superconductor
    and there is confinement.

\section{THE 'T HOOFT TENSOR} \label{tensor}
 The 't Hooft tensor can be given as an explicit gauge invariant form for any gauge
 group\cite{dlp}. This is a gauge invariant tensor equal to the residual abelian
field strength in the unitary gauge. The magnetic field coupled
to the i-th magnetic charge is that of the residual gauge group
$U(1)^{i}$ generated by $T_3^{\, i}$. The e.m. field
$A_{\mu}^{i}$ is defined in terms of the gauge field $A_{\mu}^{\prime}$ in
the unitary gauge as:
\begin{equation}
A_{\mu}^{i}=Tr(\phi^{i}_{0}\, A_{\mu}^{\prime})
\end{equation}  $\phi_{0}^i = \mu^{i}$, the fundamental weight (i = $1,\ldots,r$), identifies the monopole species.
If $b(x)$ is the transformation bringing to a generic gauge
and $A_{\mu}$ the transformed gauge field \cite{Madore1}
\begin{equation}\left\{\begin{array}{cc} A_{\mu}^{\prime} = b A_{\mu} b^{-1} - \frac{i}{g}(\partial_{\mu}b) b^{-1}\\
{}\\ \phi^i_0 = b \phi^i b^{-1}
 \end{array}\right.\label{eq.14}\end{equation}
the e.m. field is given by:
\begin{equation}
A_{\mu}^{i}=Tr( \phi^{i}( A_{\mu} + \Omega_{\mu}))
\end{equation}
where $\Omega_{\mu}= - \frac{i}{g}\, b^{-1}\partial_{\mu}b $. We
can rewrite the abelian field strength as
\begin{equation}F_{\mu\nu}^{i}=Tr(\phi^{i}\, G_{\mu \nu})+
i\, g\, Tr( \phi^{i} \, [ A_{\mu} + \Omega_{\mu}, A_{\nu}  +
\Omega_{\nu}])\label{eq55bis}\end{equation} $F^i_{\mu \nu} $ can
 computed \cite{dlp} starting
from the observation that  the ciclycity of the trace implies that
only the part of $V_{\mu}\equiv A_{\mu}+\Omega_{\mu}$, which does
not belong to the invariance group of $\phi^i$, contributes.
Indeed
\begin{equation}Tr(\phi^i[ V_{\mu}, V_{\nu}]) = Tr\, ( V_{\nu}[ \phi^i,
V_{\mu}])= Tr\, ( V_{\mu}[ V_{\nu},
\phi^i])\label{a}\end{equation} To compute the second term in
eq.(\ref{eq55bis}), it proves convenient to introduce a projector
$P^i$ on the complement of the invariance algebra of $\phi^i$, and
to write  Eq.(\ref{eq55bis}) in the form
\begin{equation}F_{\mu\nu}^{i}=Tr(\phi^{i}\, G_{\mu \nu})+
i g Tr( \phi^{i} \, [ P^i \left( A_{\mu} + \Omega_{\mu} \right),
A_{\nu}  + \Omega_{\nu}])\label{eq55}\end{equation}
It is proved in ref\cite{dlp}
 that  projection on the complement $P^i\, V_{\mu}$ is given
by
\begin{equation}P^i V_{\mu} = 1 - \prod_{\vec{\alpha}}^{\prime} \left( 1 -  \frac{ [ \phi^i,[\phi^i,\, \, ]]
}{( \vec{c}^{\, \, i} \cdot \vec{\alpha} )^2}\right) V_{\mu}\label{P}
\end{equation}
where $[ \phi^i,\, \, \,   \,  ]\, V_{\mu} = [ \phi, V_{\mu} ]$,
the product $\prod_{\vec{\alpha}}^{\prime}$ runs on the roots
$\vec{\alpha}$ such that $ \vec{c}^{\, \, i} \cdot \vec{\alpha}
\neq 0$ and only one representative of the set of the roots having
the same value of $ \vec{c}^{\, \, i} \cdot \vec{\alpha}$ is
taken.
 In order to simplify the notation we
denote by $\lambda_I^i$ the different non zero values which $(
\vec{c}^{\, \, i} \cdot \vec{\alpha} )^2$ can assume and rewrite
$P^i V_{\mu}$ as
\begin{equation}P^i V_{\mu} = 1 - \prod_{I} \left( 1 -  \frac{ [ \phi^i,[\phi^i,\, \, ]]
}{ \lambda_I^i }\right) V_{\mu}\label{P2}
\end{equation}

 By use of eq.(\ref{P2}) and recalling that
\begin{equation}D_{\mu} \phi^{\, i} = - i g [ A_{\mu} + \Omega_{\mu}\, ,\, \phi^{\, i}\, ]\end{equation}
the generalized 't Hooft tensor reads 
\begin{displaymath} F_{\mu \nu}^i = Tr ( \phi^i G_{\mu \nu} ) - \frac{i}{g}
\sum_{I} \frac{1}{\lambda_I^{i}  \,}\, \, Tr \left( \phi^i [D_{\mu} \phi^i, D_{\nu} \phi^i ] \right) +
\end{displaymath}
\begin{equation}
+\, \frac{i}{g} \sum_{I \neq J}\frac{1}{\lambda_{I}^{i}
\lambda_{J}^{i}}\, Tr \left( \phi^i [[D_{\mu} \phi^i, \phi^i],
[D_{\nu} \phi^i,\phi^i ]]\right) + \ldots \label{r1}\end{equation} In
summary, for any gauge group $G$ we have to compute for each root
$\vec{\alpha}$ the (known) commutator $[ \phi^i , E^{\vec{\alpha}}
] = ( \vec{c}^{\, \, i} \cdot \vec{\alpha} )\, E^{\vec{\alpha}}$,
where $\phi^i$ are the fundamental weights associated to each
simple root. This gives us the set of the values of
$\lambda_I^i$ to insert into eq.(\ref{r1}). For $SU(N)$ group
$[\phi^i ,E_{\vec{\alpha}}]=(  \vec{c}^{\, \, i} \cdot
\vec{\alpha} ) E_{\vec{\alpha}}$ where $(  \vec{c}^{\, \, i} \cdot
\vec{\alpha}) = 0,\pm 1$, so the projector is simply
\begin{equation} P^i V_{\mu} = [\phi^i,[\phi^i, V_{\mu} ]]\end{equation}
and the 't Hooft tensor is the usual one
\begin{equation} F_{\mu \nu}^a = Tr ( \phi^a G_{\mu \nu} ) -
\frac{i}{g}Tr ( \phi^a [D_{\mu} \phi^a, D_{\nu} \phi^a ]
)\end{equation} For a generic group the projector is more
complicated and it depends on the root chosen. For example in
$G_2$ we have two 't Hooft like tensors, one for each of the two
simple roots $e_1$ and $e_2$:
\begin{displaymath} F_{\mu \nu}^{(1)} = Tr( \phi^{(1)} G_{\mu \nu}  - \frac{5 i}{4 g}\,
  Tr\left( \phi^{(1)} [D_{\mu} \phi^{(1)}, D_{\nu} \phi^{(1)} ] \right)  \end{displaymath}\begin{displaymath}+ \frac{ i}{4 g}\, Tr \left( \phi^{(1)} [[D_{\mu} \phi^{(1)}, \phi^{(1)}], [D_{\nu}
\phi^{(1)},\phi^{(1)} ]]\right)\end{displaymath}
for the breaking of the longest simple root $e_1$ and
\begin{displaymath} F_{\mu \nu}^2 = Tr (\phi^{(2)} G_{\mu \nu}) - \frac{49 i}{36 g}\,
 Tr \left(\phi^{(2)} [D_{\mu} \phi^{(2)}, D_{\nu} \phi^{(2)} ] \right)\end{displaymath}\begin{displaymath}+
 \frac{7 i}{18 g}\, Tr \left( \phi^{(2)} [[D_{\mu} \phi^{(2)}, \phi^{(2)}], [D_{\nu}\phi^{(2)},\phi^{(2)} ]]\right)\end{displaymath}\begin{displaymath}- \frac{ i}{36 g}\, Tr \left( \phi^{(2)} [[[D_{\mu} \phi^{(2)}, \phi^{(2)}],\phi^{(2)}],[[D_{\nu}
\phi^{(2)},\phi^{(2)} ],\phi^{(2)}]]\right)
\end{displaymath}
for the breaking of the second one.
\section{DISCUSSION}

The tight experimental limits on the number of free quarks in
nature indicate that confinement is an absolute property, namely
 that the number of free quarks is strictly zero due to some
symmetry. Deconfinement is then a change of symmetry.\\
 Since color is
an exact symmetry, the only way to have an extra symmetry, to be broken, 
is to look for a dual description of QCD. The extra
degrees of freedom are infrared modes related to spatial homotopy.
This is a special case of the geometric Langlands
program of ref.\cite{Gukov}.\newline  Since the sphere at spatial
infinity has dimension 2, the relevant homotopy in 3+1 dimensions
 is  $\Pi_2$, configurations are monopoles
 and the quantum numbers magnetic
charges.\newline For a generic gauge group of rank $r$ there exist
$r$ different magnetic charges $Q^a$. The existence of nonzero magnetic
charges implies a violation of Bianchi identities by the abelian
gauge fields coupled to them. The gauge invariant abelian field
strengths coupled to $Q^a$ are known as 't Hooft tensors.

Monopoles for a generic gauge group have been analyzed and the
corresponding 't Hooft tensors computed in ref \cite{dlp}.

\end{document}